\documentclass[twocolumn,showpacs,preprintnumbers,amsmath,amssymb,floatfix]{revtex4}

\usepackage{graphicx}
\usepackage{dcolumn}
\usepackage{bm}

\begin{document}

\title{Long-Term Stability of an Area-Reversible Atom-Interferometer Sagnac
Gyroscope}

\author{D. S. Durfee}
 \affiliation{Department of Physics and Astronomy, Brigham Young University,
Provo, Utah 84602, USA}
\author{Y. K. Shaham}
\affiliation{ Department of Physics, Yale University, New Haven,
CT 06520-8120, USA }
\author{M. A. Kasevich}
\affiliation{ Department of Physics, Stanford University,
Stanford, CA 94305-4060, USA }

\date{\today}

\begin{abstract}

We report on a study of the long-term stability and absolute
accuracy of an atom interferometer gyroscope.  This study
included the implementation of an electro-optical technique to
reverse the vector area of the interferometer for reduced
systematics and a careful study of systematic phase shifts.  Our
data strongly suggests that drifts less than 96 $\mu$deg/hr are
possible after empirically removing shifts due to measured
changes in temperature, laser intensity, and several other
experimental parameters.
\end{abstract}


\maketitle

A Sagnac gyroscope measures rotation through an induced phase
shift, known as the Sagnac phase, between the two arms of an
interferometer. In recent years Sagnac gyroscopes have
demonstrated extraordinary sensitivity \cite{Gustavson00rs,
Rowe99da, Dunn02da}.  In addition to sensitivity, however, many
applications, including inertial navigation and tests of general
relativity, require an instrument with high accuracy or excellent
long-term stability \cite{Schiff60pn}.  In this paper we report
on a study of the long-term performance of an atom interferometer
gyroscope. As part of this study we implemented a method to
periodically reverse the vector area of the interferometer loop,
a technique which has already proven useful in atom interferometer
accelerometers \cite{Snadden98mo}. This reversal causes the sign
of the Sagnac phase to change, allowing systematics which do not
change sign with the reversal to be cancelled.

The apparatus has been described previously
\cite{Gustavson00rs,Gustavson00thesis}. Experimental details are
summarized here.  A pair of counter-propagating cesium atomic
beams run the length of the apparatus. The atoms are generated
thermally and have a characteristic velocity of 220 m/s
(determined from the Sagnac phase induced by Earth rotation and by
time-of-flight measurements).  The atoms are collimated by narrow
tubes, transversely laser cooled, and then optically pumped into
the $6\,^2\text{S}_{1 / 2}$ $F=3$ hyperfine ground state.  They
then interact with three pairs of counter-propagating laser beams
traveling perpendicular to the atomic beams.  The three
interaction regions are spaced by 0.968 m, for a total length
from the center of the first to the center of the last beam pair
of 1.936 m. The counter-propagating laser beams drive stimulated
Raman transitions between the $F=3$, $m_F=0$ and the $F=4$,
$m_F=0$ cesium ground-state hyperfine levels.   Laser
polarization and a magnetic bias field suppress transitions
between other magnetic sub-levels.  For the studies reported here
the two Raman lasers were tuned about 850 MHz below the
$6\,^2\text{S}_{1 / 2}$ $F=3$ and $F=4$ to $6\,^2\text{P}_{3 /
2}$ $F=3$ transitions.  This combination of parameters results in
a Sagnac phase of 9.1 radians for a rotation rate equal to Earth
rotation ($\Omega_E = 15$ deg/hr).

Each Raman transition involves the absorption of a photon from one
laser beam and the stimulated emission of a photon into the
counter-propagating beam, giving the atom two photon recoils of
transverse momentum.  As a result, the Raman lasers act as atom
``mirrors'' and ``beamsplitters'' to spatially split and then
coherently recombine the beam of atoms. In the language of Ramsey
interferometry the laser fields form a $\pi / 2 - \pi- \pi / 2$
pulse sequence. The first $\pi / 2$ pulse places the atoms into an
equal superposition of $F=3$ and $F=4$ states and gives the $F=4$
component two photon recoils of momentum.  Next the $\pi$ pulse
swaps the internal energy and external momentum states of the two
halves of the atom wave.  This causes their paths to cross at the
final $\pi / 2$ pulse which mixes them and causes them to
interfere. The atoms are then probed by resonant fluorescence to
determine the final population in the $F=4$ state.  From this
measurement the interferometer phase, and thereby the rotation
rate of the apparatus, is determined.

To implement the area reversal we modified our previous method of
generating the two Raman laser frequencies \cite{Bouyer96ms}. In
our current scheme, the +1 and -1 diffraction orders of a master
laser passing through a high frequency acousto-optic modulator
(AOM) are used to inject two slave diode lasers.  The AOM's
frequency is set such that the slave lasers differ in frequency
by the cesium hyperfine ground state splitting plus or minus a
small (5 MHz) detuning. The 5 MHz detuning from the two-photon
resonance prevents the co-propagating light frequencies from
driving recoil-free Raman transitions.  The beams from the two
slave lasers are coupled into a common fiber which delivers the
light to the interferometer.  The beam is then split into three
``Raman'' beams: two of equal intensity plus a third beam at
twice this intensity. These beams are used to generate the $\pi$
and $\pi/2$ pulses.

The sign of the 5 MHz detuning determines the direction of the
interferometer's vector area. Each Raman beam passes through the
atomic beam in one of the three interaction regions.   After
passing once through the atomic beam, each Raman beam is sent
through a pair of AOMs and then retro-reflected back through the
AOMs and the atomic beam. The double pass through the AOMs
results in a frequency upshift of 5 MHz.  This causes one of the
incoming light frequencies to form a two-photon resonance with
one of the retro-reflected frequencies. Changing the sign of the
5 MHz detuning switches which frequency components of the direct
and retro-reflected light are involved in the transition. This
causes the atoms to recoil in the opposite direction, reversing
the interferometer's area.  This electro-optic area reversal
method is extremely clean, producing a precise $180^{\circ}$
reversal and exceptional cancellation of numerous systematic
errors which change sign relative to the Sagnac phase with the
reversal.

All of the data we present was taken utilizing the
counter-propagating atomic beam and electro-optically applied
rotation bias methods described in \cite{Gustavson00rs}, allowing
us to cancel linear acceleration shifts and the Sagnac phase due
to the Earth's rotation.  The ability to bias our instrument near
zero rotation is an uncommon feature in Sagnac interferometers.
Laser gyros typically require a rotation bias to prevent
backscatter-related frequency pulling and mode locking
\cite{Stedman97rl}.  Working near zero bias makes the
interferometer insensitive to many sources of instability and
increases the fringe contrast for improved signal-to-noise ratios.
In this mode of operation the instrument's rotation rate can be
obtained by subtracting the electro-optically applied rotation
bias from the interferometrically measured rotation rate.

\begin{figure}
\includegraphics[width=8.5 cm]{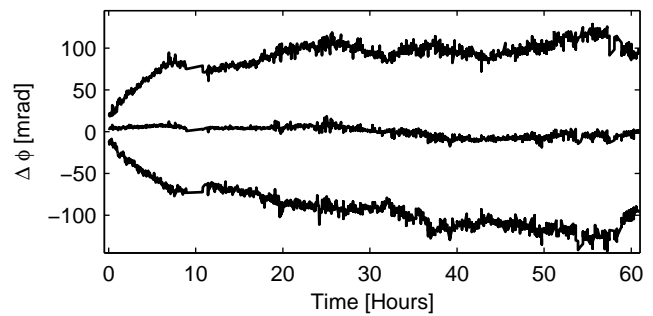}
\caption{\label{fig:longrun} Variation of measured phase over a 60
hour period. The upper and lower traces are the interferometer
phase in the area reversed and non-reversed configurations ($\pm$
a constant such that they don't overlap). The interferometer
phase of the reversed-area trace has been multiplied by -1 such
that a positive phase change indicates a positive rotation change
in both traces. The middle trace is the average of the other two
traces. Rapid drifts at the beginning of data sets, evident in
the upper and lower traces, were very common, possibly due to the
change in temperature after the experimenters left the room. Area
reversal does an extremely good job of removing this drift. For
comparison, the Sagnac phase for Earth rotation rate is 9.1
radians. }
\end{figure}

To study long-term stability we took several sets of data, each
of which span a period of about two and a half days.  A typical
data set is shown in Fig.\ \ref{fig:longrun}. Although this data
set exhibits more long-term drift than some of our other sets, it
demonstrates the cancellation of systematic drifts that can
result from area reversal. This figure presents raw data,
corrected only by the removal of a calibrated,
independently-measured phase-shift due to drifts in Raman laser
intensity.  The strong anti-correlation of the two configurations
seen in this figure, and the much lower drift in the average phase
relative to either configuration, demonstrates the utility of area
reversal for increased stability.

\begin{figure}
\includegraphics[width=8.5 cm]{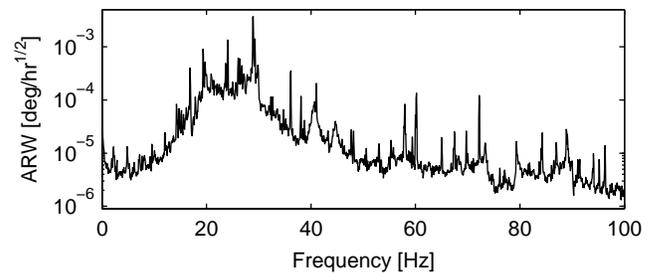}
\caption{\label{fig:arw} The power spectral density of the random
angular motion.  This plot shows the typical rotational noise
present in our laboratory.}
\end{figure}

Figure \ref{fig:longrun} shows high-frequency noise on top of
slow drifts.  The high-frequency noise is due to mechanical noise
in the building.  For measurements of long-term stability we
placed the interferometer directly on the lab floor to reduce
measurement errors due to slow drift in the position of the
device.  Although this results in efficient coupling of our
apparatus to the high-frequency mechanical noise present in the
room, for these studies we were not concerned with high-frequency
noise which quickly averages away. This noise has been well
characterized with high-speed measurements of the Sagnac phase of
our instrument.  The rotational noise, shown in Fig.\
\ref{fig:arw}, hits a baseline angle random walk (ARW) of a few
times $10^{-6}$ deg/hr$^{1/2}$ at frequencies from 2 to 7 Hz.

\begin{figure}
\includegraphics[width=8.5 cm]{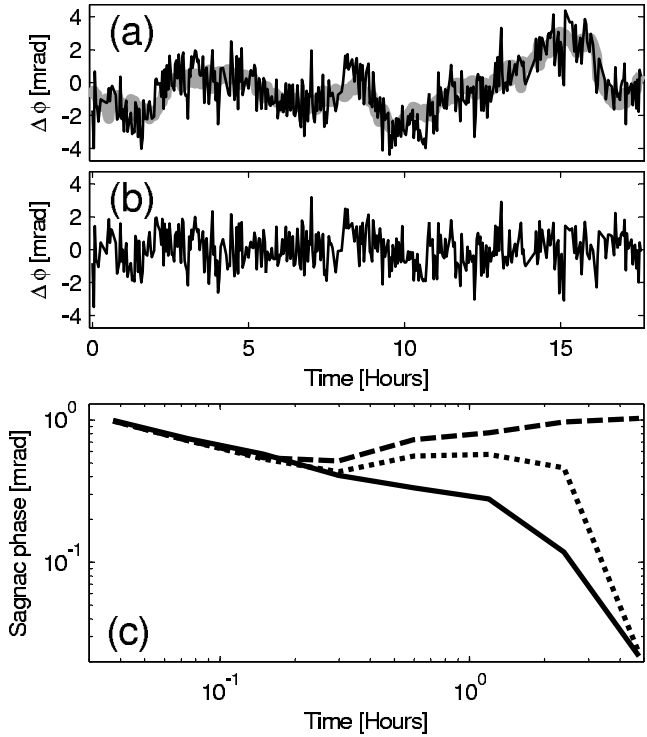}
\caption{\label{fig:correlations} The effect of removing
correlations.  The black line in (a) is the Sagnac phase from a
particularly quiet set of data plotted vs.\ time. This data
represents the average rotation phase for the area reversed and
non-reversed configuration. The thick grey line shows the
correlation of the Sagnac phase drift with 9 other measured
parameters.  The line represents the sum of the parameters each
multiplied by a fixed constant. The plot in (b) is the same
Sagnac phase data after the data is corrected by subtracting
these linear correlations. Plot (c) is the Allan deviation of the
Sagnac phase. The dashed line represents the deviation of the
uncorrected data. The solid line represents the corrected data.
The dotted line results from correcting the Sagnac phase by
removing linear correlations with just three parameters.
 }
\end{figure}

Several of our data sets showed considerably less low-frequency
drift than the data shown in Fig.\ \ref{fig:longrun} due to lower
environmental noise in the lab during these runs. One of these
data sets is shown in Fig.\ \ref{fig:correlations}(a). The Allan
variance of the raw data in this figure, shown in Fig.\
\ref{fig:correlations}(c), bottoms out at a measurement time of
about 18 minutes.  Using the methods described in \cite{Ng97co}
we calculate a bias stability of 560 $\mu$deg/hr for this data
set. This is considerably smaller than what is found from the
data in Fig.\ \ref{fig:longrun}, which bottoms out at 1.1 mdeg/hr
at a measurement time of about an hour.

The stability of most gyroscopes is improved considerably by
removing known correlations in the rotation signal with other
measured parameters such as the temperature of the apparatus. In
each experiment we measured $\sim 30$ auxiliary parameters,
including laser pointing and intensity and temperatures around the
apparatus. In quiet data sets the long-term drift shows strong
correlation with several of these parameters. A simple linear
correction, in which a subset of these quantities were each
multiplied by a constant and subtracted from the Sagnac phase at
every point in time, considerably reduced drift. The effect is
shown in Fig.\ \ref{fig:correlations}(a) and (b).  The Allan
deviation is plotted in Fig.\ \ref{fig:correlations}(c).

The Allan deviation in Fig.\ \ref{fig:correlations}(c) shows that
the corrected signal's long-term stability is limited not by
systematic error but by random noise, even for the maximum
measurement time of 4.7 hours. Because the last few points in the
Allan variance involve only a statistically small number of
clusters of data \cite{Ng97co}, rather than use the minimum Allan
deviation of 22 $\mu$rad, we calculated a more conservative
estimate of the bias stability of the corrected data by
extrapolating out to 4.7 hours the $t^{-1/2}$ behavior present in
Fig.\ \ref{fig:correlations}(c) up to measurement times of about
1 hour. This yields a minimum deviation of 88 $\mu$rad
corresponding to a bias stability of 96 $\mu$deg/hr. The character
of the Allan variance strongly suggests that the inherent
long-term drift of the apparatus could be less than this and
implies a scale factor drift of less than 7 parts per million.

In addition to stability tests, we have also re-examined
systematics in the context of area reversal
\cite{Gustavson00rs,Gustavson00thesis}.  Below we discuss only
the dominant sources.  A detailed discussion of the theory can be
found in \cite{Bongs05ho}.

\begin{figure}
\includegraphics[width=8.5 cm, height = 7.2 cm]{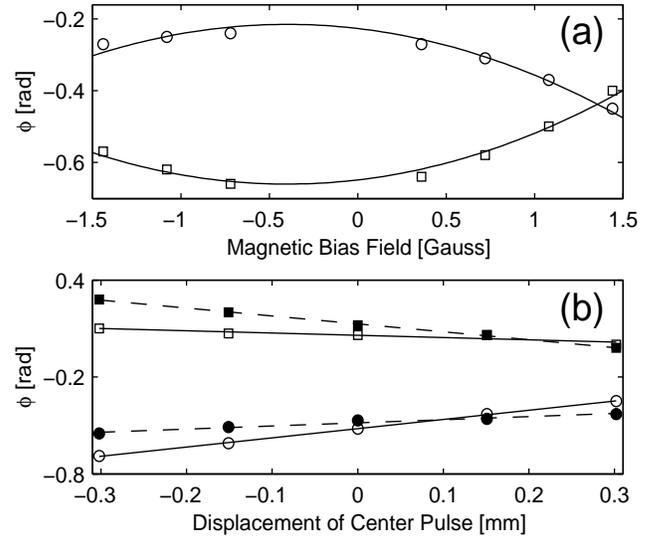}
\caption{\label{fig:magbiasandcenterbeam} Cancellation of
systematics by area reversal.  The circles/squares represent the
reversed/non-reversed configuration. The measured interferometer
phase is plotted verses (a) the applied magnetic bias field, and
(b) the center pulse displacement. The phase shift with magnetic
field in (a) is parabolic, as expected. The offset of the apex of
each parabola from zero bias indicates the presence of stray
magnetic fields. The filled and unfilled marks in (b) represent
measurements on atomic beams with two different average
horizontal transverse velocities. Averaging the reversed and
non-reversed phase cancels the magnetic bias systematic and most
of the center beam displacement systematic. }
\end{figure}

One significant error results from a $\sim 0.5$\% difference in
the magnetic bias field between the two halves of the apparatus.
The resulting quadratic Zeeman shift systematic is shown in Fig.\
\ref{fig:magbiasandcenterbeam}(a). Area reversal nearly cancels
this shift, leaving a residual shift of less than 0.1\% of
$\Omega_E$.

Another significant systematic phase shift occurs if the $\pi$
pulse is not exactly centered between the $\pi/2$ pulses.
Assuming that gravity is nearly perpendicular to the plane
defined by the interferometer and that the atomic beam is nearly
perpendicular to the Raman beams, to lowest order this
interferometer phase shift is given by
\begin{equation}
    \begin{split}
    \phi_{\Delta} = \left( \frac{\hbar {\bm{k}_{eff}}^2}{m v} +
     \frac{ 2 \bm{k}_{eff} \cdot \bm{v} }{v} +
    \frac{4 L {\bm{k}_{eff}} \cdot \left( \bm{\Omega} \times \bm{v} \right) }{v^2} + \right. \\
    \left. \frac{2 L \bm{k}_{eff} \cdot \bm{g} }{v^2} \right)
    \Delta.
    \label{eq:centerbeamdelta}
     \end{split}
\end{equation}
Here ${\bm{k}_{eff}}$ is the effective two-photon wave vector for
the Raman beams (defined such that the two-photon momentum kick
the atom receives is given by $\hbar {\bm{k}_{eff}}$) and $L$ is
the distance between pulses. The parameters ${\bm{v}}$ and $m$
are the effective velocity of the atomic beam and the atomic mass,
and ${\bm{\Omega}}$ and ${\bm{g}}$ are the rotation rate and
acceleration (gravity) present on the apparatus.  The
displacement of the $\pi$ pulse from the center of the
interferometer is represented by $\Delta$. The first term in Eq.\
\ref{eq:centerbeamdelta} is cancelled by area reversal. Given our
ability to center the $\pi$ pulse, this term can account for most
of the difference between the reversed and non-reversed phase
near zero field in Fig.\ \ref{fig:magbiasandcenterbeam}(a). The
second term (a combination of $\Delta$ and a non-zero transverse
atom velocity), and the last two terms (resulting from a
combination of inertial forces and unbalanced lengths), are not
cancelled by area reversal. We studied these systematics by
varying the pulse spacing. From the results, shown in Fig.\
\ref{fig:magbiasandcenterbeam}(b), we conclude that they
contribute an error of less than 1\% of $\Omega_E$.

Systematics due to horizontal Raman beam misalignment should be
negligible; we use observed Doppler shifts in individual
interaction regions for precision horizontal alignment. Vertical
misalignments, however, may contribute significant shifts. Tilting
the plane of the interferometer introduces a phase shift due to
gravity. This shift is cancelled using counter-propagating atom
beams \cite{Gustavson00rs,Gustavson00thesis}. Deviation of the
$\pi/2$ beams from the plane defined by the atomic beam and the
$\pi$ beam results in a gravitational shift which is not
cancelled by counter-propagating atom beams or with area
reversal. A related shift of the same magnitude results from a
combination of vertical Raman beam misalignment and a non-zero
vertical component of an atom's velocity. Intentionally walking
the vertical Raman beam alignment we could only change the
measured rotation rate by 7\% of $\Omega_E$ before signal contrast
was reduced significantly. When aligned for maximal contrast this
systematic should be below 2\% of $\Omega_E$.

Drift in Raman laser intensities introduce additional phase
shifts. The two Raman laser intensities are chosen to minimize
the differential ac-Stark shift between the $F=3$ and $F=4$
hyperfine ground-states. Because both interferometer paths
experience nearly equal ac-Stark shifts, and since Stark shifts
can be cancelled using counter-propagating atomic beams, Stark
effects should not be significant \cite{Weiss94pm,
Gustavson00rs}.  But because the pulse area received by an atom
with a given velocity depends on the intensities of the Raman
beams, the laser intensity determines which velocities result in
highest fringe contrast and contribute most to the interferometer
signal.  Since the Sagnac phase at a given rotation rate depends
on velocity, we still observe changes in phase when the
intensities of the Raman beams are varied.

Area reversal reduces intensity-dependent shifts by an order of
magnitude. Incomplete cancellation is due in part to an imperfect
implementation of area reversal; a difference in the detuning from
the single photon resonance for the two configurations and
imperfect AOM efficiency in the retro-reflections results in
slightly different pulse areas and ac-Stark shifts for the
non-reversed and reversed configurations.  Although it is
possible to improve the implementation, these shifts are not a
dominant source of instability. The difference in Sagnac phase
for the two areas at a given rotation rate is well below 1\%. And
since we typically add a rotation bias to keep the Sagnac phase
small, effects related to the atomic beam velocity which scale
the Sagnac phase without adding an independent bias phase are
greatly reduced.

In conclusion, we have implemented a method of reversing the
vector area of the interferometer to reduce systematic drifts and
have completed a study of the long-term behavior of our
gyroscope.  Removing correlations with auxiliary parameters
reveals that the intrinsic stability of our gyroscope is probably
less than 96 $\mu$deg/hr (this upper value limited by statistical
white noise) with a scale factor stability better than 7 parts
per million. We also completed an analysis of the dominant
systematics present in our apparatus and determined that the
absolute accuracy is better than a few percent of Earth rotation.

We acknowledge helpful discussions with J.M. McGuirk, J.B. Fixler,
G.T. Foster and G. Nogues.  This work was supported by the
National Science Foundation, the Office of Naval Research, and
NASA.

\newpage


\end{document}